\documentclass[conference]{IEEEtran}
\usepackage[letterpaper, left=.62in, right=.62in, bottom=1.1in, top=0.75in]{geometry}

\usepackage{cite}
\usepackage{amsmath,amssymb,amsfonts}
\DeclareMathOperator*{\argmin}{arg\,min}

\usepackage{algorithmic}
\usepackage{graphicx}
\usepackage{textcomp}
\usepackage{xcolor}
\usepackage[caption=false, font=footnotesize]{subfig}

\begin{document}

\title{Quadratic approximate dynamic programming for scheduling water resources: a case study
}

\makeatletter
\newcommand{\linebreakand}{%
  \end{@IEEEauthorhalign}
  \hfill\mbox{}\par
  \mbox{}\hfill\begin{@IEEEauthorhalign}
}
\makeatother

\author{\IEEEauthorblockN{Agustin Castellano}
\IEEEauthorblockA{\textit{Universidad de la República} \\
Montevideo, Uruguay\\
acastellano@fing.edu.uy}
\and
\IEEEauthorblockN{Camila Martínez}
\IEEEauthorblockA{\textit{Universidad ORT Uruguay} \\
Montevideo, Uruguay\\
martinez\_c@ort.edu.uy}
\and
\IEEEauthorblockN{Pablo Monzón}
\IEEEauthorblockA{\textit{Universidad de la República} \\
Montevideo, Uruguay\\
monzon@fing.edu.uy}
\linebreakand
\IEEEauthorblockN{Juan Andrés Bazerque}
\IEEEauthorblockA{\textit{Universidad de la República} \\
Montevideo, Uruguay\\
jbazerque@fing.edu.uy}
\and
\IEEEauthorblockN{Andrés Ferragut}
\IEEEauthorblockA{\textit{Universidad ORT Uruguay} \\
Montevideo, Uruguay\\
ferragut@ort.edu.uy}
\and
\IEEEauthorblockN{Fernando Paganini}
\IEEEauthorblockA{\textit{Universidad ORT Uruguay} \\
Montevideo, Uruguay\\
paganini@ort.edu.uy}
}

\maketitle

\begin{abstract}
We address the problem of scheduling water resources in a power system via approximate dynamic programming. To this goal, we model  a finite horizon economic dispatch problem with convex stage cost and affine dynamics, and consider a  quadratic approximation of the  value functions. Evaluating the achieved policy entails solving a quadratic program at each time step, while value function fitting can be cast as a semidefinite program. We test our proposed algorithm on a simplified version of the Uruguayan power system, achieving a four percent cost reduction with respect to the myopic policy.
\end{abstract}


\section{Introduction}

Optimal operation of multi-reservoir systems for economic dispatch is a topic that has been extensively studied \cite{yeh, labadie, rani2010simulation}. Succintly, the goal is to obtain a sequence of release decisions that achieve system operation with minimal cost over a planned horizon, while also meeting operational constraints. In systems involving large reservoirs decisions become coupled across time, while also being dependant on the availability of water ---which is typically stochastic. The usual framework for solving these kinds of problems is (Stochastic) Dynamic Programming, where the state of the system typically includes the storage level in each reservoir. Standard practice involves discretizing the state variable and computing the value function at each point. However, the number of needed evaluations grows exponentially with the number of states, a phenomenon known as the Curse of Dimensionality \cite{powell}. In order to circumvent this issue, several (approximate) techniques have risen which allow for the problem to be solved in continuous spaces. One of such celebrated algorithms is SDDP which seeks to approximate the value function by a set of lower bounding affine functions \cite{pereira}. However, getting a rich enough approximation might entail the use of too many hyperplanes \cite{porteiro}. Moreover, under quadratic stage cost and affine dynamics the resulting value functions are provably convex quadratic \cite{boydOtro}. Given this, we sought to explore an alternative simpler parametric model. Specifically, we aim to tackle this problem by approximating each value function with a suitable convex quadratic function. This simplified model allows us to formulate the scheduling problem as a special case of convex approximate dynamic programming, therefore making the problem tractable on a continuous state manifold while also relaxing the need of computing exact averages, something typical of SDDP \cite{pereira}.

There exist a vast literature on approximate convex dynamic programming. For a certain class of scalar storage problems, the value functions can be proven to be convex piecewise linear, and algorithms with proven convergence guarantees have been developed \cite{nascimento_powell_13, nascimento_powell_09}. Quadratic approximate dynamic programming has been used before (see e.g. \cite{boydOtro}), especially for systems with quadratic cost and transition dynamics that are affine in the control (see \cite{keshavarz2014quadratic} for further examples including trajectory tracking and portfolio optimization). We build on these contributions for modelling the water scheduling problem as a quadratic approximate dynamic program.

The paper is outlined as follows. Section \ref{sec:DP} introduces our dynamic programming model with inflow evolutions. Section \ref{sec:Alg} presents the proposed algorithm, which involves sequentially solving several quadratic programs \cite[p.152]{boyd} and one semidefinite program \cite{boydSDP}. In section \ref{sec:Test} we present our numerical results applied on the Uruguayan power system, while also detailing how to incorporate hydrologic uncertainty in our model. Conclusions are given in Section \ref{sec:Conclu}.

\section{Hydroelectric System Modelling}\label{sec:DP}

Consider a  model of operation of a hydroelectric system over a horizon $K$, with time indexed as $k=0,1,\ldots,K-1$. A state vector $x_k\in\mathbb{R}^n$ represents current storage level at $n$ reservoirs; a control vector $u_k\in\mathbb{R}^m$ models the actions taken by the system operator, including the release and spill term on each hydroelectric plant.
Water inflows $w_k\in\mathbb{R}^p$ at a subset of the reservoirs are modeled as correlated noise. Notice that we do not enforce $p=n$ since there might not be significant inflows at some of the reservoirs. The cost of operation of the system is modeled through a function $g_k(x_k,u_k,w_k)$, which may include the cost of thermal generation and a penalty for deviating from economic dispatch. Our goal is to obtain a sequence of control actions $\mathbf{u}=\{u_0,\ldots,u_{K-1}\}$ such that, for a given starting state $x$, the expected cost of running the system is minimized:


\begin{align}
    & \mathbb{E}_{w_0}\left[\min_{u_0}g_0(x_0, u_0, w_0)+\mathbb{E}_{w_1}\left[\min_{u_1}g_1(x_1, u_1,w_1)+\ldots\right.\right.\label{eq:DP0}\\
    &+\left.\left.\mathbb{E}_{w_{K-1}}\left[\min_{u_{K-1}}g_{K-1}(x_{K-1}, u_{K-1},w_{K-1})\right]\right]\mid x_0 = x\right]\nonumber\\
    &x_{k+1} = f_k(x_k,u_k,w_k)\nonumber
\end{align}

Notice that this formulation ---which first computes a minimum and then an expected value--- differs from typical dynamic programming approaches \cite{bertsekas}, where the order is inverted. Implicitly, we are assuming at the $k-$th stage that the disturbance $w_k$ is known. This means full knowledge of total inflows at the start of each time interval.

Dynamic Programming allows for decoupling of the optimization problem \eqref{eq:DP0} across stages. For this purpose, let us define the cost-to-go function from stage $k$ onwards:

\begin{align}
    &V_k(x) =  \mathbb{E}_{w_k}\left[\min_{u_k}g_k(x_k, u_k, w_k)\right.\\
    &+\mathbb{E}_{w_{k+1}}\left[\min_{u_{k+1}}g_{k+1}(x_{k+1}, u_{k+1},w_{k+1})+\ldots\right.\nonumber\\
    &+\left.\left.\mathbb{E}_{w_{K-1}}\left[\min_{u_{K-1}}g_{K-1}(x_{K-1}, u_{K-1},w_{K-1})\right]\right]\mid x_k = x\right]\nonumber\\
    &x_{\kappa+1} = f_\kappa(x_\kappa,u_\kappa,w_\kappa)\nonumber
\end{align}


As usual, the main idea behind this decoupling is to compute the cost-to-go for stage $k+1$ and, in a recursive manner,  use this solution to compute the cost-to-go for stage $k$ by using Bellman Equation \cite{bertsekas}:

\begin{equation}
    V_k(x_k) = \mathbb{E}\left[\min_{u_k\in \mathcal{U}_k(x_k,w_k)}\left\{g_k(x_k,u_k,w_k)+V_{k+1}(x_{k+1})\right\}\right]\label{eq:BellmanEquality}
\end{equation}

where we explicited box constrains $\mathcal{U}_k(x_k,w_k)$ depending on the current state and inflow (e.g.: release and spill terms must be non-negative and bounded), power balance, etc. It can be shown that the value functions are convex, given that the stage cost is convex and the transition dynamics are affine in both the state and the control \cite{keshavarz2014quadratic}.

\subsection{Hydrologic state space model}

To capture correlations in water inflows across stages we expand the state variable to include a discrete Markov state $e_k=e$ that summarizes the current hydrological environment. Its dynamics are governed by an homogeneous Markov chain, with transition probabilities:

\begin{equation}
    \mathcal{P}_{ee'} = {P}\left(e_{k+1}=e'\mid e_k=e\right)
    \label{eq:MarkovTransition}
\end{equation}

This probabilities may be estimated from historical data. One possibility is letting $e_k$ take two values (corresponding to dry and wet) as introduced in \cite{philpott}.
Local practice in Uruguay is to use a 5-level model which spans from very-dry to very-wet \cite{casaravilla}, with transitions given by a non-homogeneous Markov chain. We propose keeping this 5-level discretization while modelling the hydrologic state evolution as time invariant. This entails procuring a single transition matrix $\mathcal{P}\in\mathbb{R}^{5\times 5}$ from the available data, which will be accomplished using Principal Component Analysis \cite{PCA1, PCA2}. A more thorough description of our proposed model is presented later in Section \ref{sec:MarkovEstimation}.

We separate the hydrologic state $e$ from the reservoir levels $x$ and solve the  expected value in Bellman Equation in two steps. Since this hydrologic state can only take discrete values, we can compute a different value function $V_{k,e}(x)$ for each possible value of $e$. Then, for given $e_k=e$, we estimate the future cost-to-go by an expected value over the next hydrologic state $e'$, computed according to the finite probabilistic model given in \eqref{eq:MarkovTransition}. The generalized Bellman iteration thus becomes:


\begin{align}
V_{k,e}(x) = \mathbb{E}&\left[\min_u\left\{ g_k\left(x, u, w\right)+\sum_{e'}\mathcal{P}_{ee'}.V_{k+1,e'}\left(x'\right)\right\}\right]\label{eq:BellmanExact}\\
&x'=f_k(x,u,w)\nonumber\\
&{u \in \mathcal{U}_{k}\left(x,w\right)}\nonumber
\end{align}


where the outmost expectation is taken over inflows $w$ conditioned to $e_k=e$. The rightmost sum in \eqref{eq:BellmanExact} can be interpreted as an estimate of the future cost-to-go given the current hydrologic state. If the costs $g_k$ and the dynamics $f_k$ are affine \eqref{eq:BellmanExact} is a linearly constrained quadratic program \cite[p.152]{boyd} and can be efficiently solved using standard techniques.



\section{Algorithm}\label{sec:Alg}

\subsection{Backward pass}

As has been argued before, our goal is to compute approximate value functions $\tilde{V}_{k,e}(x)$ quadratic in $x$, for every stage $k$ and hydrologic state $e$. Each iteration of the backward dynamic programming algorithm is subdivided into two parts: a \textit{sampling stage}  and a \textit{fitting stage}. The sampling stage consists of obtaining state-cost pairs $(x,\beta)$ by solving an approximate Montecarlo-based version of \eqref{eq:BellmanExact}:

\begin{align}
\hat{\beta}_{k,e}(x,w_i) &= \min _{u}\left\{ g_k\left(x, u, w_i\right)+\sum_{e'}\mathcal{P}_{ee'}.\tilde{V}_{k+1,e'}\left(x'\right)\right\}\label{eq:BellmanInexact}\\
\beta_{k,e}(x)&=\frac{1}{M}\sum_{i=0}^{M-1}\hat{\beta}_{k,e}(x,w_i)\label{eq:averageBeta}
\end{align}
Upon obtaining $N$ pairs $\left(x_k^{s},\beta_{k,e}^{s}\right)$, we fit the quadratic value function by solving:

\begin{align}
\min_{P, q, r} &\sum_{s=0}^{N-1} \left({x_k^s}^\top P x_k^s +q^\top x_k^s +r - \beta_{k,e}^s\right)^2 \label{eq:fit_quadratic}\\
\text{s. t.: }&P \succeq 0\nonumber
\end{align}

The computational complexity of our proposed method resides in solving $N\times M$ linearly constrained quadratic programs (as in \eqref{eq:BellmanInexact}) and one semidefinite program (as in \eqref{eq:fit_quadratic}) for each stage and hydrologic state.

\subsection{Forward pass}\label{sec:fwdpass}

Once all the value functions are approximated, the expected cost of running the system from a certain initial state $x$ and certain hydrologic state $e$ could be obtained by evaluating the fitted function $\tilde{V}_{0,e}(x)$. However, each stage of the backwards phase introduces errors on the approximations, and therefore the predicted cost $\tilde{V}_{0,e}$ might differ from the true cost substantially. In order to gauge the actual cost obtained by our methodology, a forward phase is carried out. This phase implements a Montecarlo simulation scheme which sequentially solves the one-stage optimization problem:

\begin{align}
u_k =&\argmin_{u \in \mathcal{U}_{k}\left(x_k,w_k\right)}\left\{ g_k\left(x_k, u, w_k\right)+\sum_{e'}\mathcal{P}_{e_ke'}.\tilde{V}_{k+1,e'}\left(x_{k+1}\right)\right\}\label{eq:ForwardPass}\\
&x_{k+1}=f_k(x_k,u,w_k)\nonumber
\end{align}

starting at $k=0$ with initial storage level $x_0=x$ and  hydrological state $e_0=e$. 
The incurred cost of operation over the planned horizon is the expected sum of the running cost per stages:

\begin{equation}total\;cost (x, e) = \mathbb{E}\left[ \sum_{k=0}^{K-1}g_k(x_k,u_k,w_k)\mid x_0 = x, e_0=e\right]\label{eq:TotalCost}\end{equation}


where the expectation is taken over all possible sequences $\left\{(w_k, e_{k+1})\right\}_{k=0}^{k=K-1}$, and the control laws $u_k$ are derived from \eqref{eq:ForwardPass}. This simulated cost corresponds to deploying our policy, and is therefore a better figure of merit for evaluating performance than the predictions $\tilde{V}_{0,e}(x)$.

Moreover, the obtained policy's performance can be contrasted with the performance of the $\textit{myopic policy}$, which at time $k$ seeks to minimize the current stage cost:

\begin{align}
u_k^{\textit{myopic}} =&\argmin_{u \in \mathcal{U}_{k}\left(x_k,w_k\right)} g_k\left(x_k, u, w_k\right)\label{eq:ForwardPassMyopic}\\
&x_{k+1}=f_k(x_k,u,w_k)\nonumber
\end{align}

Intuitively, at each step the myopic policy will use up (possibly all) the available water, minimizing the current cost and disregarding the utility of water in the future. While at first glance a reasonable thing to do, this behavior is generally suboptimal due to the expected inflows over the next steps and the spatial interconnection of the dams. For example, it could be better suited to store water now (at the expense of a higher cost) for use later, when a drought is expected.

We expect our methodology to outperform the myopic policy. But how good can our policy really be? Although this question remains unanswered, we can construct a \textit{lower bound} on the optimal performance. For a \textit{given} inflow sequence $\mathbf{w}=\{w_0,\ldots,w_{K-1}\}$ the optimal decisions $\mathbf{u}=\{u_0,\ldots,u_{K-1}\}$ and the optimal cost can be obtained by solving the $K-$stage problem:


\begin{align}
    \mathbf{u}^{LB}(x,\mathbf{w}) &= \argmin_{\mathbf{u}\in\mathbf{U}(x,\mathbf{w})} \sum_{k=0}^{K-1}g_k(x_k,u_k,w_k)\label{eq:LowerBound}\\
    &x_{k+1}=f_k(x_k,u_k,w_k)\quad \forall k=0,\ldots,K-1\nonumber\\
    &x_{0}=x\nonumber
\end{align}

where $\mathbf{u}\in\mathbf{U}(x,\mathbf{w})$ means that $u_k\in\mathcal{U}_k(x_k,w_k)$ for each $k$, with the sets $\mathcal{U}_k(x_k,w_k)$ described in \eqref{eq:BellmanEquality}. Problem \eqref{eq:LowerBound} solves for the whole decision sequence $\mathbf{u}=\{u_0,\ldots,u_{K-1}\}$ at once, by being given full knowledge of all the noise realizations $\mathbf{w}$ at the start of the planning horizon. This is in sharp contrast with our proposed algorithm, where at each stage $k$ the controller only has access to the current noise $w_k$. The expected cost of running \eqref{eq:LowerBound} over all the possible inflow sequences $\mathbf{w}$ is indeed a lower bound on \eqref{eq:TotalCost} since the expectation of the minimum is lower than the minimum of the expectation. Intuitively, \eqref{eq:LowerBound} ahcieves a lower value because more information about the future inflows is available for planning.



\section{Test case: the Uruguayan system}\label{sec:Test}
\subsection{The Uruguayan system}

Uruguay is a small country with a demand profile that seldom surpasses $2000MW$. It is comprised of $4$ hydroelectric plants: $3$ of which are located in a cascade-like fashion along the \textit{R\'io Negro} basin; the fourth one is located in the \textit{R\'io Uruguay}, and is shared with neighbouring Argentina. The combined installed power in said facilities is roughly $1500MW$. There are a number of wind farms in Uruguay, with a total installed power amounting to more than 75\% of the country's peak load. In recent years, there has been a surge in the installation of solar farms as well \cite{adme}.

We will employ a one-year horizon with weekly decisions ($K=52$ weeks in a year, $k=0,\ldots,K-1$). In that regard, non-dispatchable renewables (wind and solar) will be left out of our model since they typically vary on a much faster timescale. Generation will be provided by the four hydroelectric plants and by a single thermal generator representing the aggregate thermal generation of the whole system. The state vector $x_k\in\mathbb{R}^4$ represents the current volume at each of the four reservoirs. The control $u_k=\left[r_k^\top, s_k^\top, t_k\right]^\top\in\mathbb{R}_{+}^9$ consists of the release ($r_k\in\mathbb{R}_{+}^4$) and spill vectors ($s_k\in\mathbb{R}_{+}^4$) and the total thermal generation ($t_k\in\mathbb{R}_{+}$). The state dynamics are described by

\begin{equation}
    x_{k+1} =f(x_k,u_k,w_k)= x_k + B \left(r_k+s_k\right) +w_k
\end{equation}

where $B$ is the coupling matrix that captures the interconnection between hydro plants:

\begin{equation}
B=\left(\begin{array}{cccc}-1 & 0 & 0 & 0 \\ 1 & -1 & 0 & 0 \\ 0 & 1 & -1 & 0 \\ 0 & 0 & 0 & -1\end{array}\right)
\end{equation}

and the vector $w_k$ gathers the weekly inflows at each reservoir, as detailed in the next Section. Finally, the cost function $g(t_k)$ is the cost incurred by thermal generation, modeled as linear and time-invariant.

\subsection{Markov Model estimation}\label{sec:MarkovEstimation}

The series used in this case study consist of the weekly measured inflows from the three main reservoirs in Uruguay collected over 105 years (1905--2009). As a first step, we clean up the negative values which correspond mainly to measurement errors, and for this study are considered as Not Aviailable (NA) data in the model estimation phase. We then proceed through several steps:

\subsubsection{Normalization}
Each one of the three series of hydraulic inflow is divided by its weekly median across the time period to remove the seasonal variations along the year. The second step is to apply a logarithm transformation to the normalized series (Box-Cox transformation with $\lambda=1$ \cite{shumway2017time}). After these two normalization steps, it can be observed that the new series present an approximately Gaussian distribution. In Figure \ref{fig:medianas_histogramas} we plot the estimated median inflow and the resulting distribution after transformation.

\begin{figure}
    \centering
    \includegraphics[width=0.8\columnwidth]{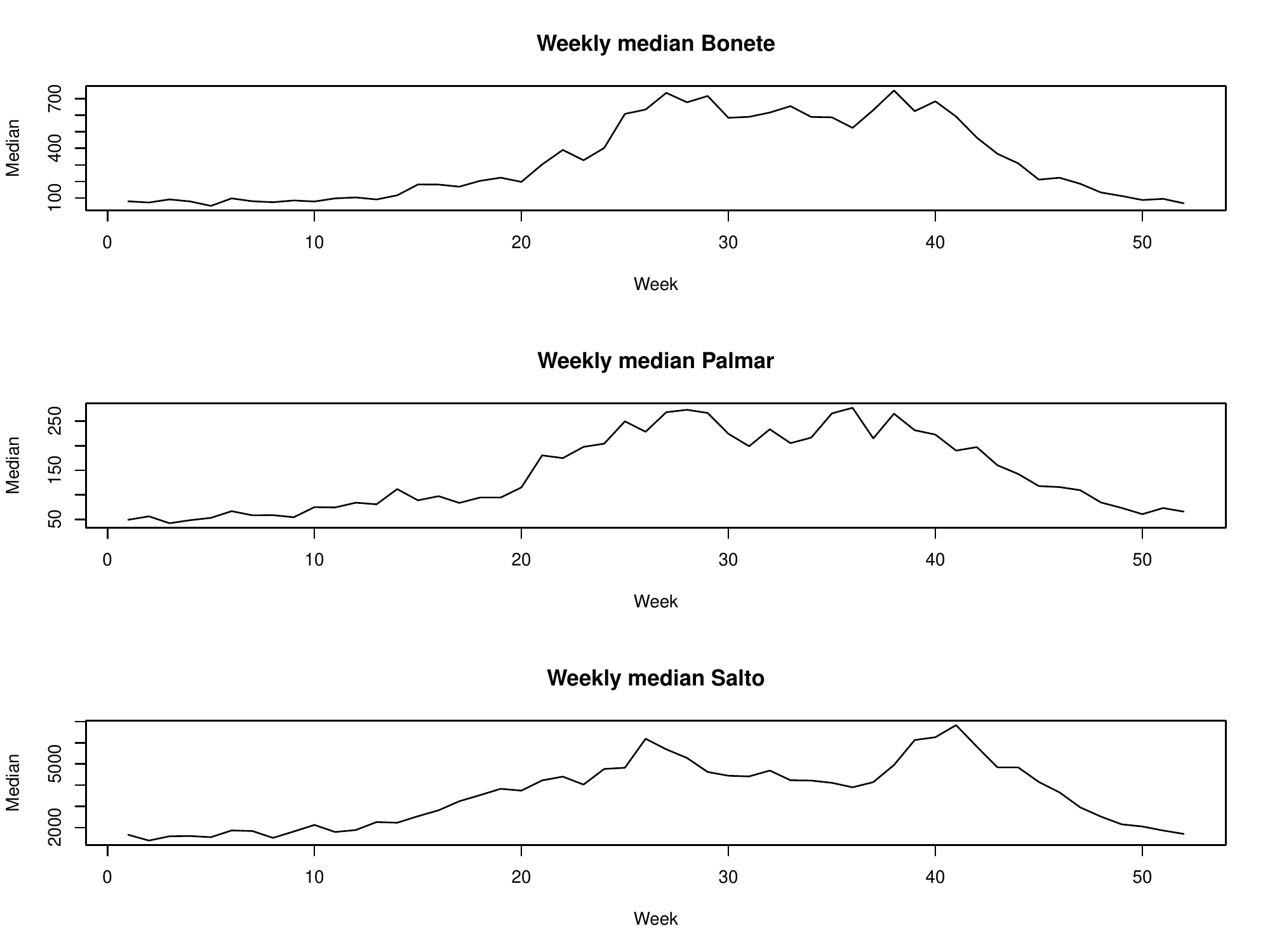}\\
    \includegraphics[width=0.8\columnwidth]{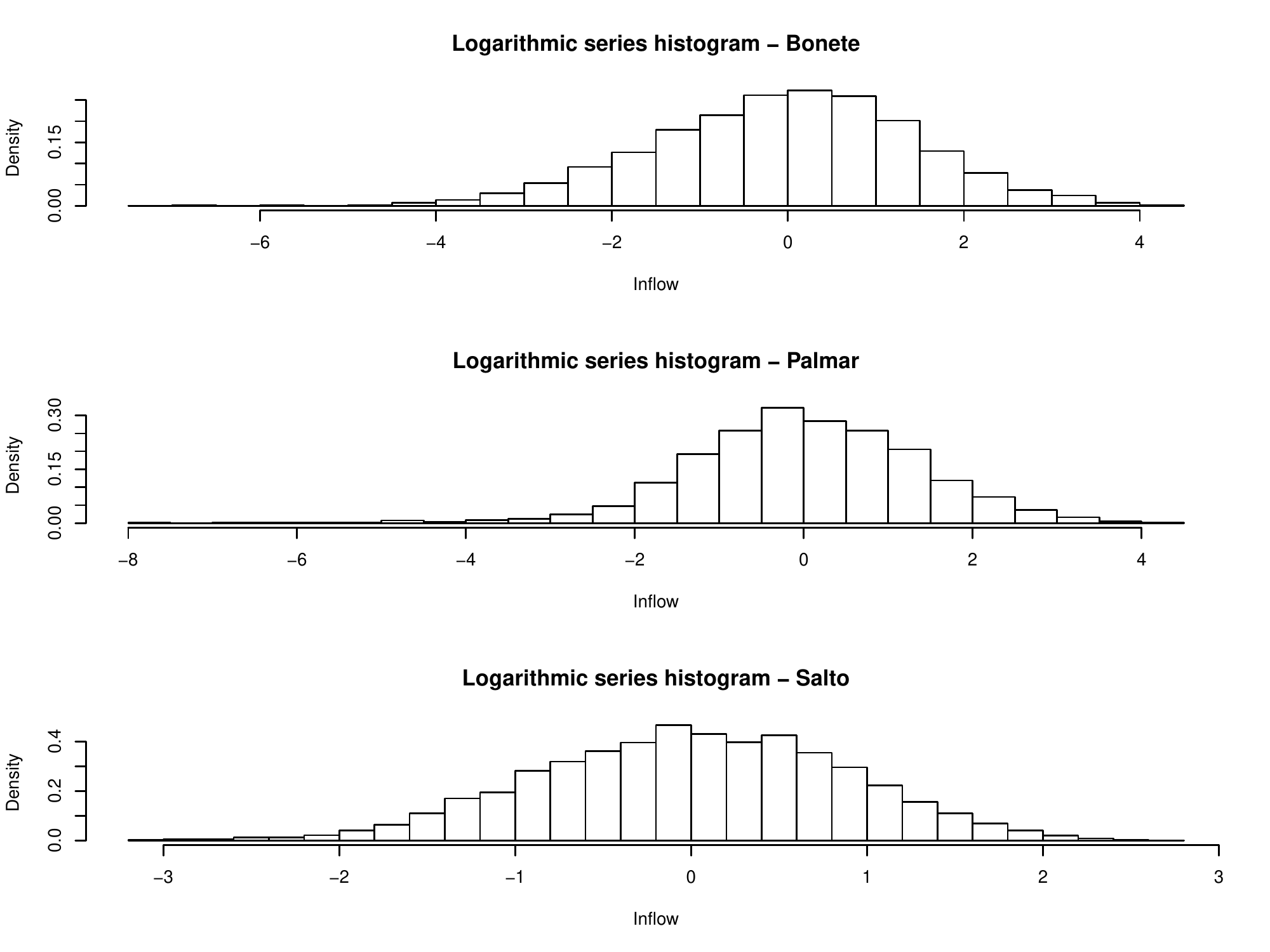}
    \caption{Observed median inflow for the three reservoirs and histogram of the normalized logarithmic inflows.}
    \label{fig:medianas_histogramas}
\end{figure}

\subsubsection{Model estimation}
As mentioned before, in this phase entire rows with NA data are removed. To adjust a Markov model to the 
historic inflows, we considered two clustering techniques in order to group similar inflows in a fixed amount of categories. Our first approach was to perform a $K$-means algorithm \cite{shumway2017time} applied directly in the three dimensional space of log inflows. This algorithm consists in separating the data into $K$ clusters in a way that the euclidean distance between each point to the centroid of the assigned group is minimized. Given an initial but not optimal clustering, the algorithm relocates each point to its new nearest center, update the clustering centers by calculating the mean of the updated members, and repeat the relocating-and-updating process until convergence criteria (such as predefined number of iterations, difference on the value of the distortion function) are satisfied. In our study we considered a $K=5$ clusters, obtaining the clusterization depicted in Figure \ref{fig:clusters}.

As a final step, the rows not assigned to any cluster due to NA values (which came from zeros in the original dataset) are replaced by a small value (in order for the logarithmic transformation to work) and labelled accordingly. That is, they are not used to fit the clusters, but are labelled using the clusters obtained with the previous data. This ensures a more robust estimation of the clusters.

\begin{figure}
    \centering
    \includegraphics[width=0.9\columnwidth]{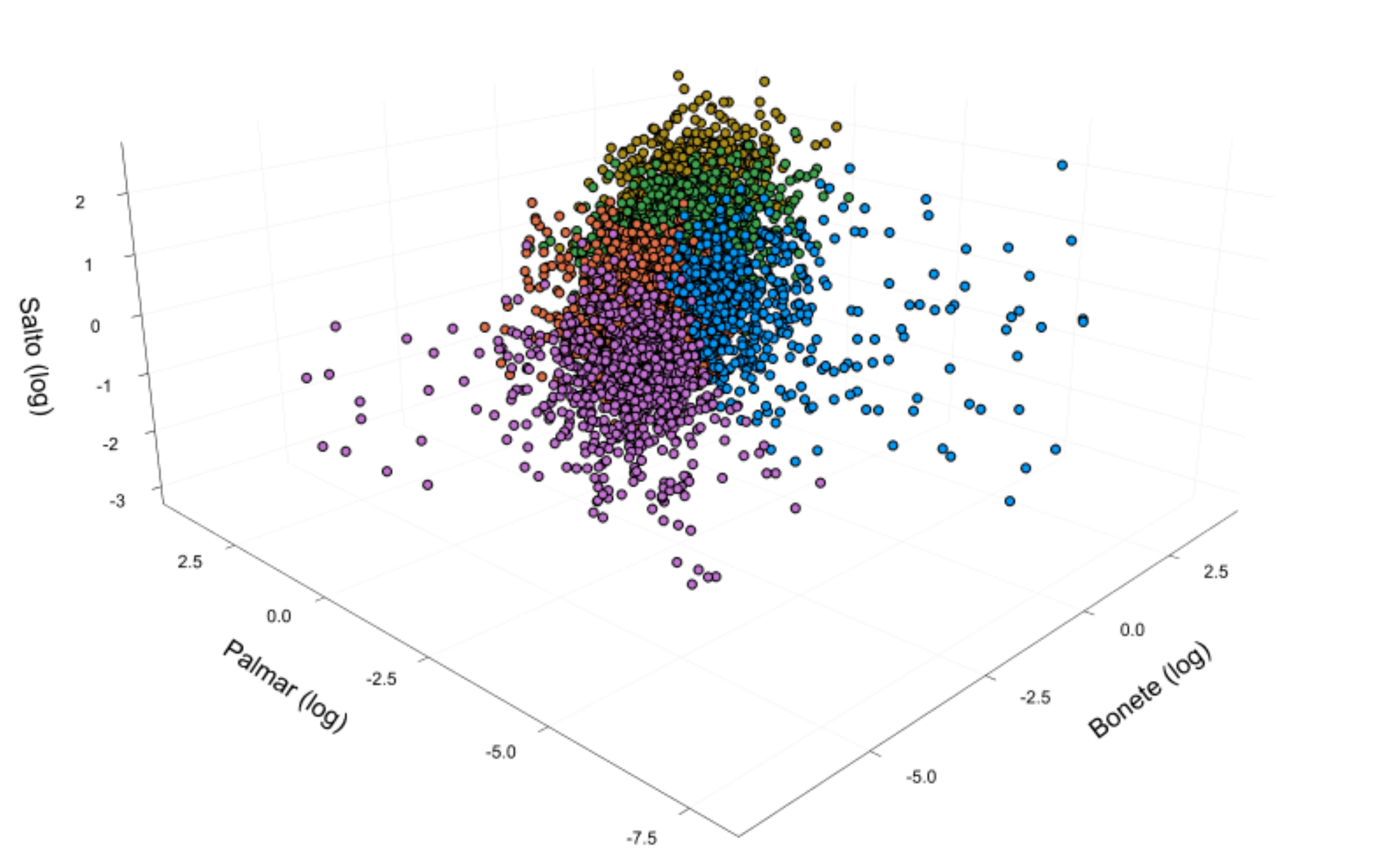}\\
    \includegraphics[width=0.9\columnwidth]{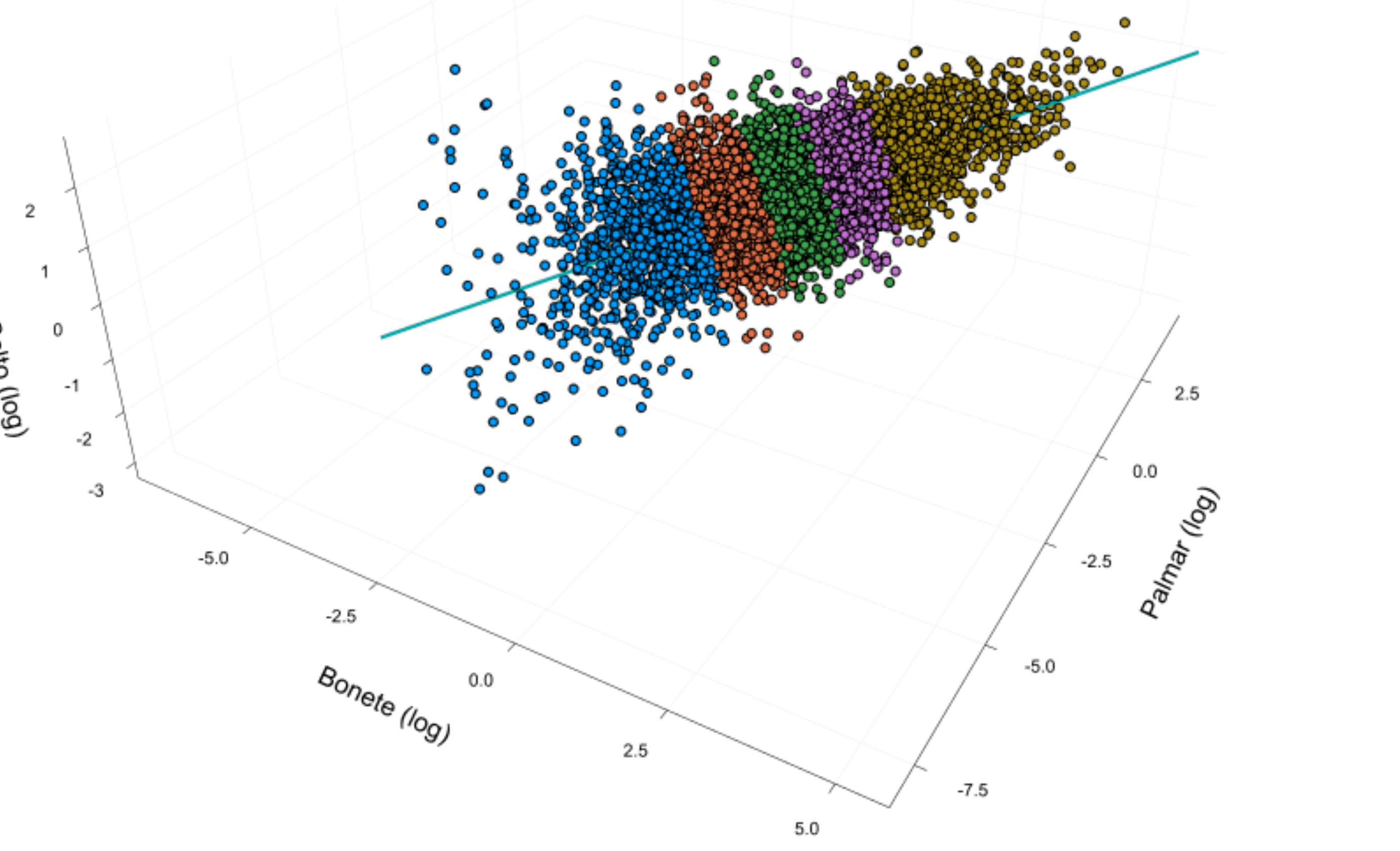}
    \caption{Obtained clusters for the $K-$means approach and PCA approaches.}
    \label{fig:clusters}
\end{figure}

The problem with this approach is that the cluster code $1,\ldots,5$ is not related to the hydraulicity of the group because the centroids are assigned randomly by the algorithm. This is why we considered a second approach where the clustering is applied to a suitable \emph{feature} of the inflow vectors. We chose to construct this feature using principal component analysis (PCA) by projecting the points along the principal component axis of the inflow vector. This ensures that the feature variability is maximized.

To find the principal direction, the covariance matrix of the data set is calculated, as well as its eigenvectors and eigenvalues. The principal direction is the one associated with the largest eigenvalue, and before doing the projection into this direction, the eigenvector is normalized so the projection is just a convex combination of inflows (i.e. the sum of the components of the PCA eigenvector are normalized to sum $1$). After projection, we chose $K=5$ clusters containing $20\%$ of the observations each, with increasing inflows in each category. The resulting clusters now have physical meaning, with larger inflows being integrated in the same cluster, and they are depicted in Figure \ref{fig:clusters}.

\paragraph{Markov process estimation}\hfill

For both clustering methods explained before, in order to estimate the Markov process, the transitions between different clusters are counted. 

The parameters $p_{ij}$ of the markovian matrix $\mathcal{P}$ represent the probability of making a transition from cluster $i$ to cluster $j$. This parameters are calculated as follows:

\begin{equation}
   \widehat{p}_{ij} = \frac{\sum_{t=1}^{T}\mathbf{1}_{e_{t-1=i}, e_{t=j}}}{\sum_{t=1}^{T}\mathbf{1}_{e_{t=i}}}
\label{eq:pij}\end{equation}
where $e_t$ represents the state in time $t$, and the sum is computed along the complete state sequence.

\subsection{Simulation using Markov transitions}
Trials begin at an initial state $x_0$ and initial hydrologic state $e_0$. At each time step the hydrologic sequence is updated with the markovian matrix $\mathcal{P}$ derived in \eqref{eq:pij}, and a disturbance vector $w_k$ corresponding to said hydrologic state is sampled. In order to approximate the total cost of running the system, we substitute the expected value in \eqref{eq:TotalCost} with a sample mean carried out over $T=105$ forward passes.

\subsubsection{Performance for varying training points}

State-cost pairs are sampled by partinioning the state space in a grid-like fashion. Each of the four reservoirs $i=0,\ldots,3$ is uniformly partitioned in $N_i$ steps, yielding a total number of  $N=N_0\times N_1\times N_2\times N_3$ state points. The cost at each point is obtained by averaging over $M=10$ different noise realizations. It is worth emphasizing that the state variables are not discretized, but these grid points are knots where we anchor our quadratic model to find the specified parameters using \eqref{eq:fit_quadratic}. The results shown herafter are for varying $N_0$, which corresponds to the discretization of the largest reservoir \textit{Bonete}. For the other reservoirs we fix $N_i=3$. As an illustrating example, Fig. \ref{fig:quad_fit} shows a cut of the quadratic obtained for the fourtieth week of the year with $N_0=10$.

\begin{figure}[t]
    \centering
  \subfloat{%
       \includegraphics[width=0.55\linewidth]{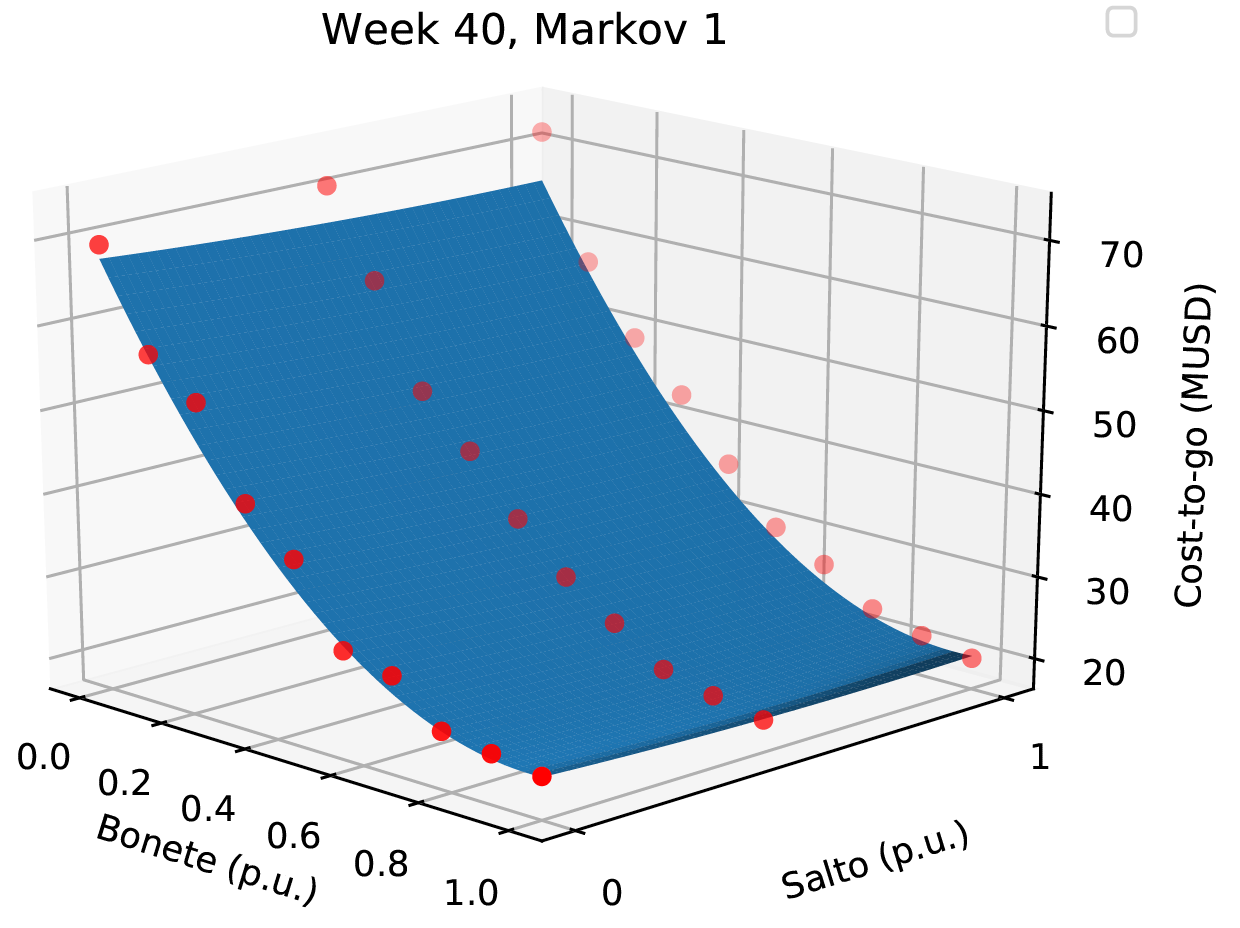}}
    \hfill
  \subfloat{%
        \includegraphics[width=0.45\linewidth]{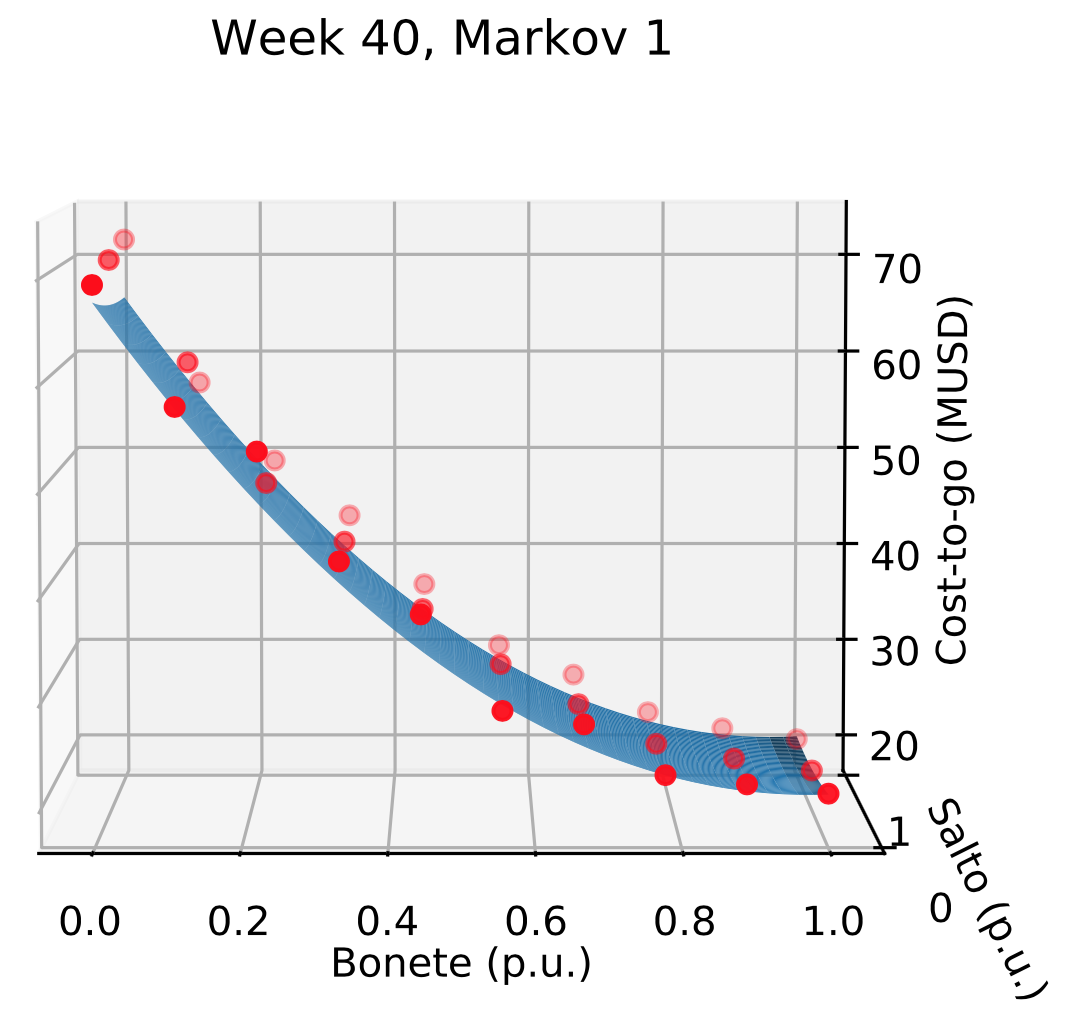}}
  \caption{Fitted quadratic function for the fourtieth week of the year and wet hydrologic state $(e_{40}=1)$ for $N_0=10$. In red: sampled state-cost pairs (using \eqref{eq:BellmanInexact}--\eqref{eq:averageBeta}). In blue: fitted quadratic function (using \eqref{eq:fit_quadratic}). Note that the cost-to-go seems to primarily depend on the state of $\textit{Bonete}$ dam.}
  \label{fig:quad_fit} 
\end{figure}

Sampling more state-cost pairs at every stage naturally increases the computational effort required to perform the backward pass. Nonetheless, our experiments show that there is no significant performance gain in the obtained policy if more points are used in the training phase (Fig. \ref{fig:comparison}).

    

\begin{figure}[h!]
    \centering
    \includegraphics[width=\linewidth]{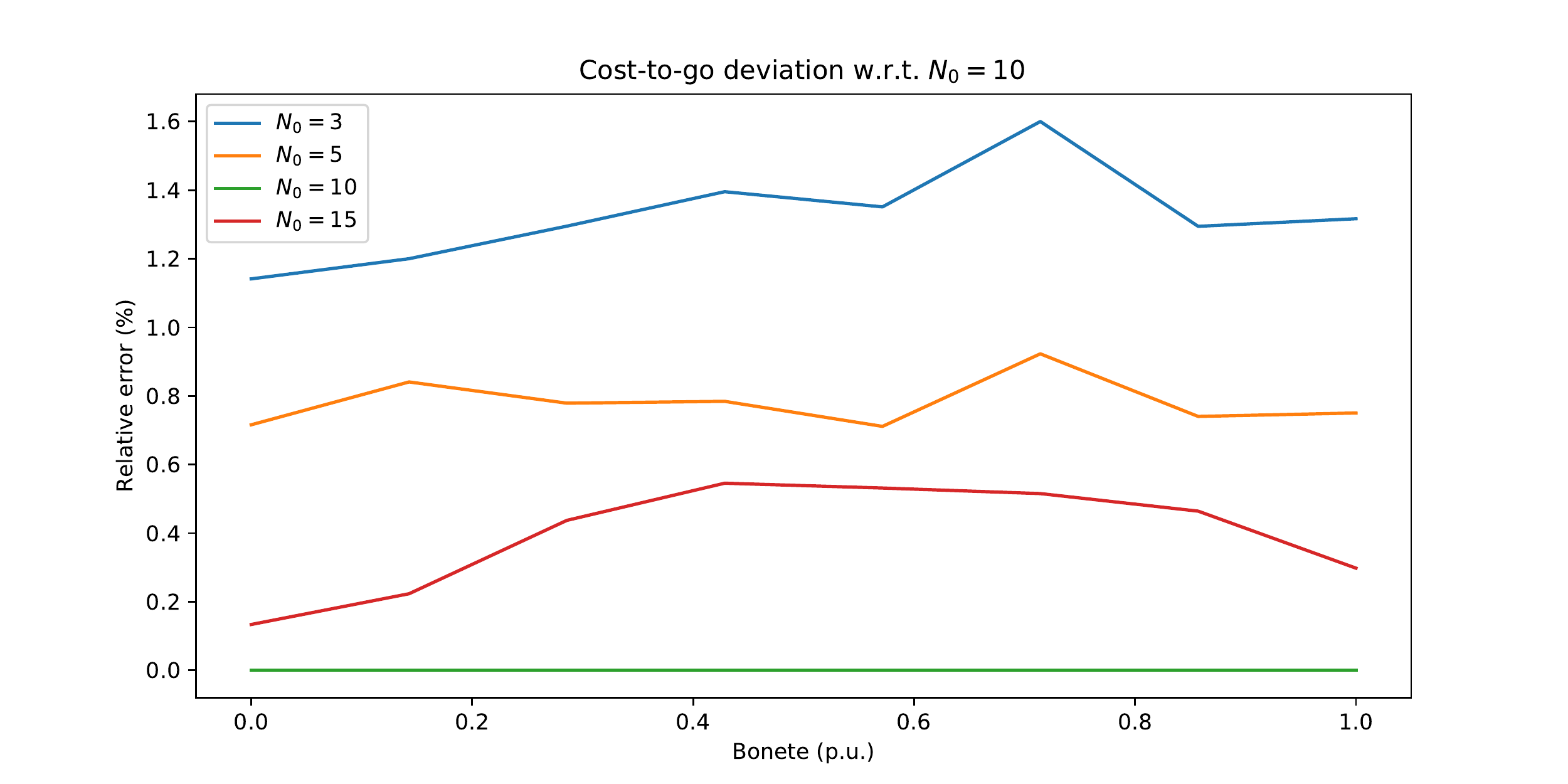}
    \caption{Percentual cost deviation with respect to $N_0=10$  as a function of Bonete's initial level. Each line corresponds to a different policy trained with varying degree of discretization of Bonete ($N_0$). Note that all the policies attain a similar cost.
    }
    \label{fig:comparison}
\end{figure}

\subsubsection{Bounds on performance and comparison with myopic policy}

We can compare the predicted cost-to-go at the start of the year $V_{0,e_0}(x_0)$ with the simulated total cost $\sum_{k=0}^{K-1}g_k$ achieved by running the system forward starting from $e_0$ and $x_0$, following the learned policy (see \eqref{eq:ForwardPass}--\eqref{eq:TotalCost}). Fig. \ref{fig:mkv0} shows a comparison between the predicted and simulated cost as a function of the level of the largest reservoir, while starting from a neither-dry-nor-wet hydrologic state ($e_0=2$). A lower bound is constructed by solving the $K-$stages problem \eqref{eq:LowerBound} given full knowledge of the noise realizations. Our experiments show that the predictions $\tilde{V}_{0,e}$ are typically optimistic.

\begin{figure}[h]
    \centering
    \includegraphics[width=\linewidth]{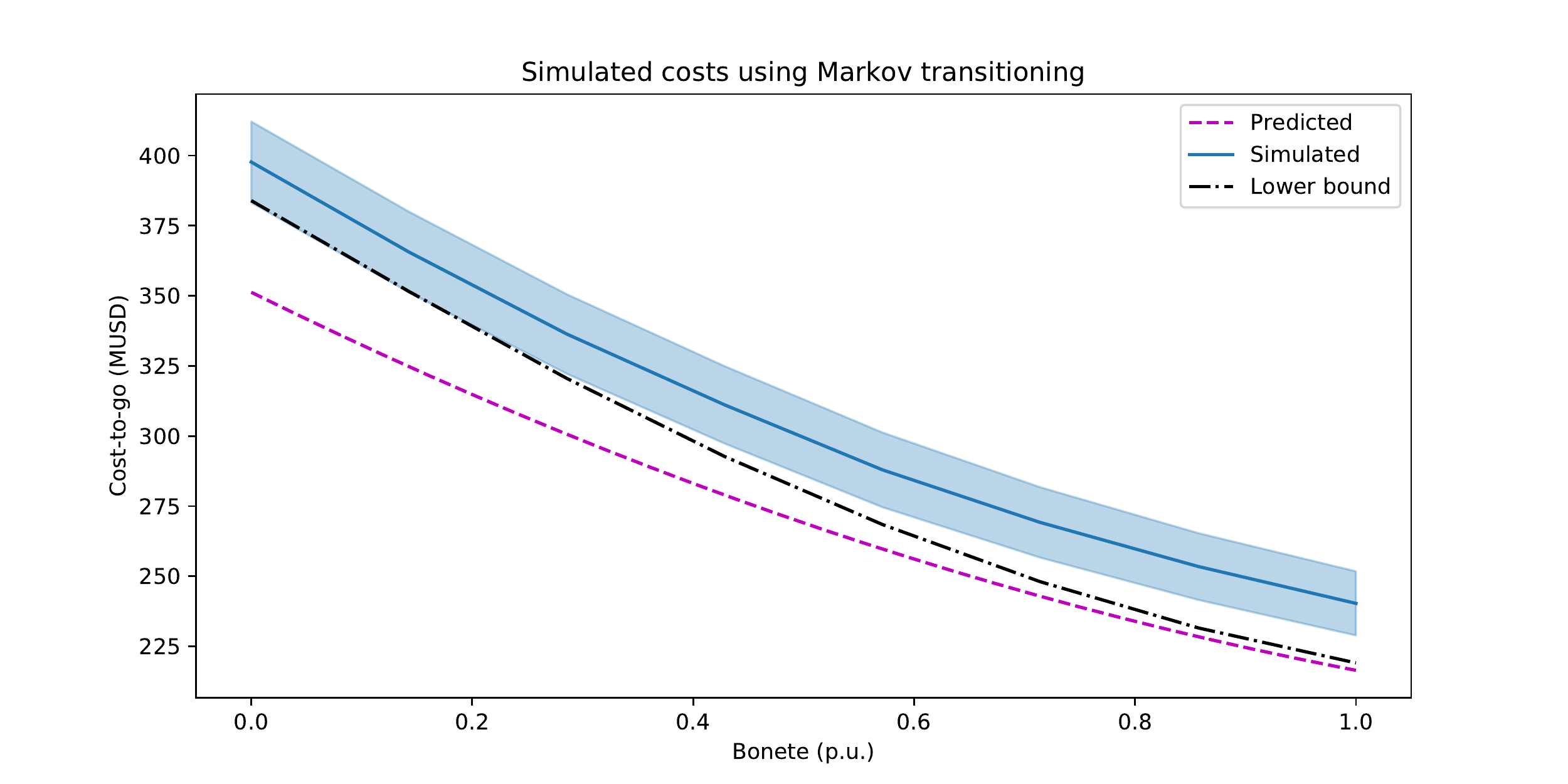}
    \caption{Annual cost as a function of largest reservoir initial level for $e_0=2$: comparison between simulated and predicted costs, along with a performance bound. Simulated costs are averaged over $T=105$ different trials. Mean cost is plotted in solid blue; shaded interval is defined as $\pm\sqrt{\sigma}/{T}$ where $\sigma$ is the sample deviation.}
    \label{fig:mkv0}
\end{figure}

The policy achieved by our proposed algorithm typically outperforms the so-called \textit{myopic policy} \eqref{eq:ForwardPassMyopic}, in particular for non-empty initial reservoir levels, as portrayed in Fig. \ref{fig:mkv_costs}.

\begin{figure}[h]
    \centering
    \includegraphics[width=\linewidth]{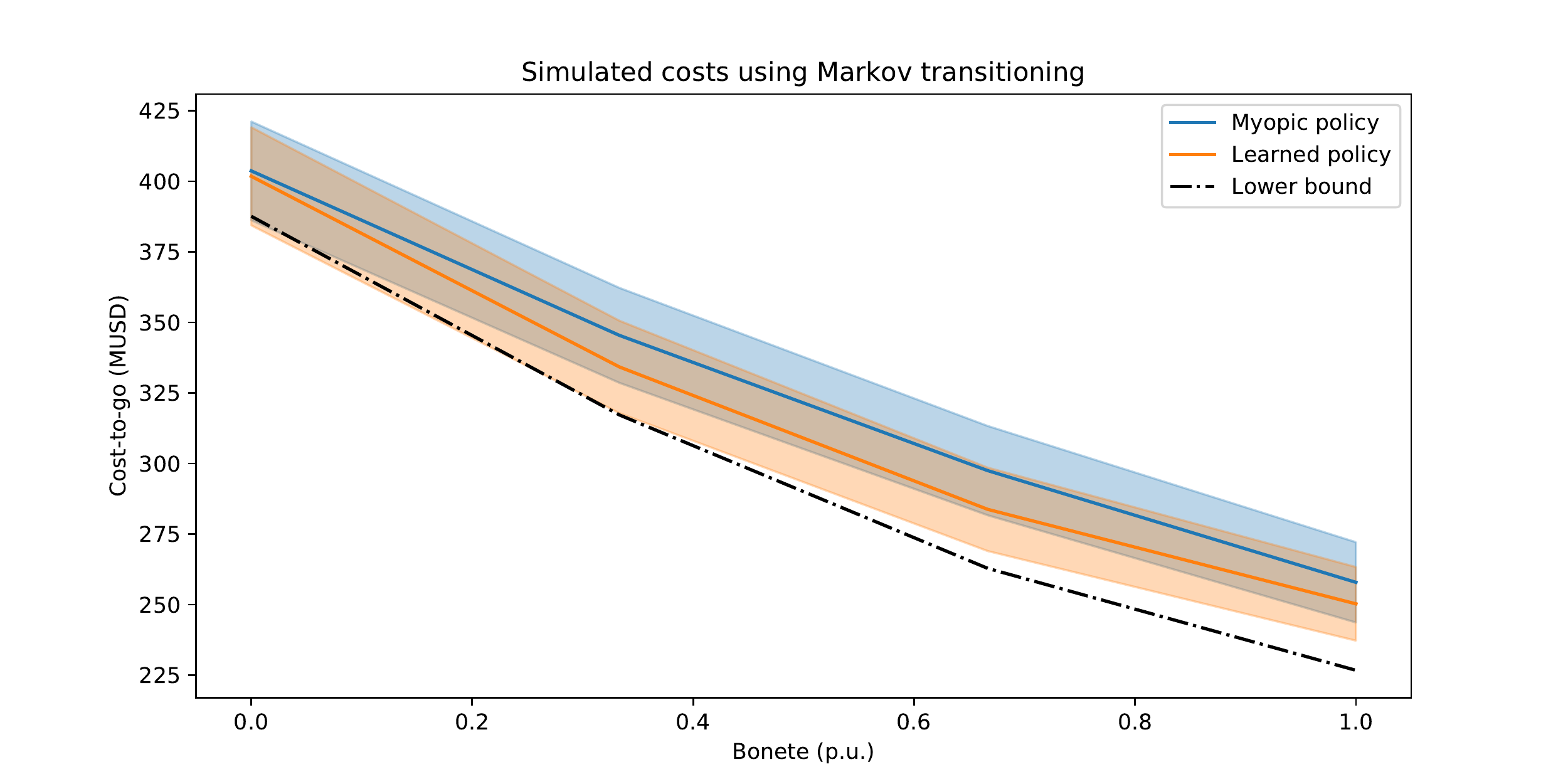}
    \caption{Comparison between the myopic (blue) and learned policy (orange) for varying initial storage levels, along with a performance bound. Our policy achieves a $4\%$ reduction on cost w.r.t. the myopic policy when storage levels are half-full, and performs at most $9\%$ worse than the lower bound policy.}
    \label{fig:mkv_costs}
\end{figure}


\subsection{Simulation using historical series}
We also perform simulations using the historical series of inflows that were used for fitting our markov model. We compare the cost attained by our policy with the cost attained by a policy that was trained with the Markov model currently in use in Uruguay, and obtain better performance (see Fig. \ref{fig:hist_todas}).

\begin{figure}[t]
    \centering
    \includegraphics[width=\linewidth]{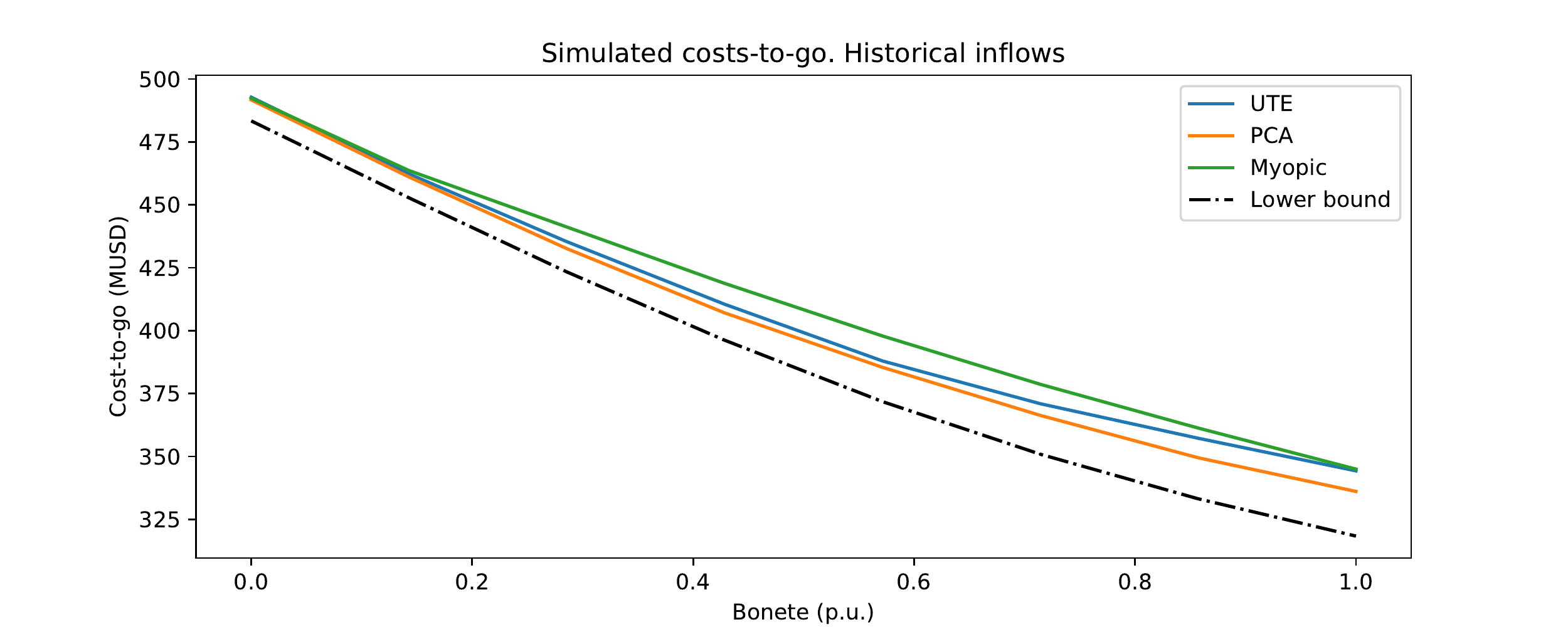}
    \caption{Mean annual cost as a function of largest reservoir (Bonete). Inflow sequences correspond with historical data. The policy trained with our proposed Markovian model (PCA) achieves a $2.4\%$ reduction on cost w.r.t. a policy trained with the model currently used in Uruguay (UTE).
    }
    \label{fig:hist_todas}
\end{figure}



\section{Conclusions}\label{sec:Conclu}
We proposed the use of convex quadratic functions to approximate the cost-to-go of a simple economic dispatch problem. We showed that training our method involves solving a sequence of  quadratic and semidefinite programs, which can be done with standard convex suites. We benchmarked our algorithm on the Uruguayan power system, obtaining performance that surpasses that of a myopic policy by four percent, and comparable to the  theoretical lower bound derived in Section \ref{sec:fwdpass}.

\section*{Acknowledgements}

This work was supported by UTE under Project UTE-FJR-UdelaR-ORT PT 001 2018 and ANII Uruguay under project FSE\_1\_2018\_1\_153050.

\bibliographystyle{./bibliography/IEEEtran}
\bibliography{./bibliography/refs}

\end{document}